\documentclass[12pt]{article}
\usepackage{amssymb}
\usepackage{amsmath}
\usepackage{epsfig}
\usepackage{graphicx}
\usepackage{color}
\usepackage{ulem}

\makeatletter

\@addtoreset{equation}{section}
\makeatother
\newcommand{\be}{\begin{equation}}
\newcommand{\ee}{\end{equation}}
\newcommand{\ben}{\begin{equation*}}
\newcommand{\een}{\end{equation*}}
\newcommand{\bea}{\begin{eqnarray}}
\newcommand{\eea}{\end{eqnarray}}
\newcommand{\alg}{\begin{align}}
\newcommand{\algx}{\end{align}}

\begin{document}

\begin{titlepage}
\vskip1cm
\begin{flushright}
\end{flushright}
\vskip1.25cm
\centerline{
\bf \large Geometric Monte Carlo and Black Janus Geometries} 
\vskip1cm \centerline{ 
Dongsu Bak,$^{\, \tt a,e}$  Chanju Kim,$^{\, \tt b}$ Kyung Kiu Kim,$^{\, \tt c}$ Hyunsoo Min,$^{\, \tt a}$ Jeong-Pil Song$^{\, \tt d}$ }
\vspace{1cm} 
\centerline{\sl  a) Physics Department,
University of Seoul, Seoul 02504 \rm KOREA}
 \vskip0.3cm
 \centerline{\sl b) Department of Physics, Ewha Womans University,
  Seoul 03760 \rm KOREA}
 \vskip0.3cm
 \centerline{\sl c) Department of Physics,~College of Science, Yonsei University, Seoul 03722 \rm KOREA}
 \vskip0.3cm

 \centerline{\sl d) Department of Chemistry, Brown University, Providence, Rhode Island 02912 \rm USA}
 \vskip0.3cm
 
  \centerline{\sl e)
B.W. Lee Center for Fields, Gravity \& Strings}
 \centerline{\sl 
 Institute for Basic Sciences, Daejeon 34047 \rm KOREA}
\vskip0.3cm
 
 \centerline{
\tt{(\,dsbak@uos.ac.kr,~cjkim@ewha.ac.kr,~kimkyungkiu@gmail.com,}} 
 \centerline{
 \tt{hsmin@uos.ac.kr,~jeong\_pil\_song@brown.edu}\,)}
  \vspace{1cm}

\centerline{ABSTRACT} \vspace{0.75cm} \noindent
{ 
We describe an application of  the Monte Carlo method to the Janus deformation of 
the black brane background. 
We present numerical results for three and five dimensional black Janus geometries with planar and 
spherical interfaces. 
In particular, we argue that the 5D geometry with  a spherical interface has an 
application in understanding the finite temperature  bag-like QCD model via the AdS/CFT correspondence.  
The accuracy and convergence of the algorithm are evaluated with respect to the grid spacing. 
The systematic errors of the method are determined using an exact solution of 3D black Janus.
This numerical approach for solving linear problems is unaffected initial guess of a trial solution
and can handle an arbitrary geometry under various boundary conditions in the presence of source fields. 
}
\end{titlepage}



\section{Introduction}

In this note, we shall consider various black Janus geometries numerically  in three and five dimensional spaces. Janus geometries are dual to interface
(conformal) field theories \cite{Bak:2003jk,Clark:2004sb}\footnote{ For a recent discussion of Janus systems, see
 \cite{Bak:2016rpn},
  where one may find a rather comprehensive list of references on the subject.}, 
 which are well-controlled deformations of the AdS/CFT correspondence \cite{Maldacena:1997re}.
  A black Janus geometry is dual to the finite temperature version of 
the corresponding interface (conformal) field theory. 
While an exact solution for the 3D black Janus geometry is available \cite{Bak:2011ga,Bak:2013uaa}, we shall  numerically reconsider it for a  
geometric interpretation of Monte Carlo (MC) method. 
In five dimensions, we shall consider two cases: one with a planar interface and the other with a spherical interface. In the latter, the boundary value of the scalar field, 
whose exponential is 
corresponding to  
the Yang-Mills (YM) coupling squared divided by  $4\pi$, has a smaller value  inside  the sphere
than  outside. Its dual field theory, whose finite-temperature counterpart shall be considered below, resembles the MIT bag model in QCD \cite{Chodos:1974je}.

 As for the numerical analysis, we shall use the standard MC method \cite{MC01} 
(see \cite{MC02} for a general review) to solve the scalar field equations in 
the black brane background which are elliptic partial differential equations (PDEs).
The choice of the MC method is conceptually motivated by the following considerations.
In the MC method {for a PDE}, an estimate to the solution at {each} site is evaluated with an average of samples of boundary values 
by generating a sufficiently large number of random walks {each of which starts from the original site and ends at one of the boundary sites.} 
The direction {of the movement in each step of the random walk}
will be chosen randomly with  respect to the probabilities determined by the {associated} PDE. 
{As we will see,} the probabilities at each site fully reflect the underlying geometry {and hence} these random walks
{may be regarded as processes exploring} geometric landscapes rather efficiently.
 This feature coins the name of ``geometric MC method'' and is the reason why this MC method {provides an interesting framework}
 for the numerical study of gravity problems. Adding to this,
  the independence between random walks allows high performance parallel computing  to speed up the convergence of the MC simulations.
Along with improved computational capabilities,
this MC method has {some} advantages compared to other numerical methods.
One can use our MC method 
on gravitational problems with arbitrary geometries with various boundary conditions.
This approach for solving linearized equations does not require any trial configuration.
It can be easily extended to full nonlinear problems of gravity theories by combination with an 
iterative scheme.

This paper is organized as follows. In sec. 2, we describe a theoretical background
for the linearized black Janus. Numerical details of our geometric MC method are
discussed in sec. 3. In sec. 4, we present our numerical analysis of 5D black Janus with planar and spherical interfaces, followed by
concluding remarks in sec. 5.

\section{Black Janus deformations to the leading order}\label{Sec2}

In this note, we shall consider the Einstein-scalar system with a negative cosmological constant described by the action
\be
I= -\frac{1}{16\pi G}\int d^{d} x \sqrt{g} \left[ R -g^{ab} \partial_a \phi \partial_b \phi + \frac{(d-1)(d-2)}{\ell^2} \right]
\label{einstein}
\ee
where $\ell$ is the AdS radius scale which we shall set to be unity for our numerical study below. For $d=3$ and $5$, this system can be 
consistently embedded into the type IIB supergravity and, hence, via the AdS/CFT correspondence, microscopic understanding of dual CFT$_{d-1}$
can be given \cite{Bak:2003jk}. In particular, in five dimensions, the dual CFT$_4$ is identified 
with the well known ${\cal N}=4$ SU(N) super-Yang-Mills (SYM) theory 
 \cite{Maldacena:1997re}.
The scalar field originated from the dilaton of the type IIB supergravity is dual to the Lagrange-density operator of the 
SYM theory, 
whose boundary value corresponds to  the logarithm  of the YM coupling squared divided by $4\pi$ in the field-theory side.  

The finite-temperature black brane background is given by
\be\label{bbmetric}
ds^2 = \frac{1}{z^2} \left[ 
(1-z^{d-1}) d \tau^2 + \frac{dz^2}{1-z^{d-1}} + dx^2_1 + dx^2_2 +\cdots + dx^2_{d-2}
\right]
\ee
with a trivial scalar field $\phi=\phi_0$. By requiring the regularity of geometry around $z=1$ in ($\tau$, $z$) space, the period of $\tau$-direction angle variable
can be identified, whose inverse is  the Gibbons-Hawking temperature of the boundary system,
$
T=(d-1)/(4\pi \ell)$.
This black brane background is dual to the finite-temperature version of CFT$_{d-1}$ on $\mathbb{R} \times \mathbb{R}^{d-2}$. The temperature may be 
scaled to other values by appropriate scaling transformations but it plays a role of unique reference scale in this pure black brane background.  

In this work, we shall consider various Janus deformations of the above black brane background. The Janus deformation in the bulk involves a scalar field 
whose boundary values jump from one to another across an interface. The dual boundary system is described by an interface CFT where its original CFT is 
deformed by an exactly marginal operator which is dual to the bulk scalar field. From the viewpoint of the boundary, its coupling jumps across the interface
from one value to another whose detailed identification is subject to the standard dictionary of the AdS/CFT correspondence.      
In $d=3$, to the leading order of the deformation parameter, the profile of the scalar field  is governed by
\be\label{3dEq}
(1-X^2) \partial_X \Bigl[ (1-X^2) \partial_X \phi \Bigr] + 4p(1-p)\partial^2_p \phi -4 p \partial_p \phi =0
\ee
where we introduce new coordinates ($X$, $p$) by $X= \tanh x_1$ and $p=z^2$. Here we shall consider the case of a single interface
which is located at $x_1= X=0$. Since the constant solution $\phi=\phi_0$ can be added freely, the Janus boundary condition can be given as
\be
\phi(X,0) = \gamma \, \, {\rm sign}({X})
\ee
where $\gamma$ is our deformation parameter referred to as an  `interface coefficient'. Of course one may consider the case of multiple interfaces \cite{Bak:2013uaa} but here we would like to focus 
on the case of a single interface. Since the leading order  is linear, we shall omit the $\gamma$ dependence 
for the simplicity of our presentation. Of course 
the validity of our approximation requires $\gamma \ll 1$ and our numerical result for the scalar profile should be understood with an extra 
multiplication  factor of $\gamma$ throughout this note. The boundary condition at $X=\pm 1$ then becomes $\phi(\pm 1, p )=\pm 1$. On the horizon side, 
one may impose  the `{Neumann} boundary condition'
$\bigl[ \sqrt{1-p}\, \partial_p \phi  (X, p)\bigr]\big|_{p=1} =0$. Note that the ($\tau$, $p$) plane of the black brane geometry has a shape of an infinite sized disk
whose center is located at $p=1$. Near this center $1-p \sim 0$, the distance from the center is approximately given by
{$s \sim \sqrt{1-p}$.} Then the above boundary condition follows {from} the Neumann boundary condition $\partial_s \phi\big|_{s=0} =0$ with respect to the
distance $s$, which ensures the smoothness of our scalar profile at $s=0$. Below we shall replace this boundary condition by a smoothness 
condition of the scalar field at $p=1$ which basically allows us to Taylor-expand $\phi(X, p)$ around $p=1$ to some orders, 
whose details  will be further specified in 
our numerical study below. We shall refer to this as a `free boundary condition'.  

Now note that our system possesses a $\mathbb{Z}_2$ symmetry $\phi(X, p)= - \phi(-X, p)$. So the problem can be reduced to solving the differential equation 
restricted in the region of $X \ge 0$ with the boundary condition at $X=0$ specified by $\phi(0, p)=0$. In this $d=3$ case, an exact solution can be found as \cite{Bak:2011ga}
\be\label{exact3Dsol}
\phi= \gamma \frac{X}{\sqrt{X^2+p(1-X^2)}}
\ee 
Even an analytic  black Janus solution including the full gravitational back-reaction has been found in \cite{Bak:2011ga}. Thus this 3D problem 
will serve as a nice testing ground for the methods we use for our numerical study below.

Now let us turn to our main theme  which is the $d=5$ case. This is relevant to the problem of understanding properties 
of ${\cal N}=4$ SYM theory. Especially in its finite temperature version, it has been argued to be useful in understanding certain aspects of the real-world 
QCD although its full justification is not that straightforward \cite{Bak:2007fk}.  

Again in the probe limit, the 5D scalar equation is reduced to 
\be
\bigl( \partial^2_{x_1}+ \partial^2_{x_2}+ \partial^2_{x_3}\bigr) \phi  + 4p(1-p^2)\partial^2_p \phi -4(1+ p^2) \partial_p \phi =0
\ee
where $p=z^2$ as before. We shall first study a Janus deformation involving a single planar interface located at 
$x_1=0$ which has  translational symmetries  along $x_2$ and $x_3$ directions. Thus  with $\partial_{x_2} \phi =\partial_{x_3} \phi=0$, the scalar profile 
is governed by
\be\label{5Deq}
(1-X^2) \partial_X \Bigl[ (1-X^2) \partial_X \phi \Bigr]  + 4p(1-p^2)\partial^2_p \phi -4(1+ p^2) \partial_p \phi =0
\ee
where we introduce $X$ by $X= \tanh x_1$ as before. For this planar interface, we have the boundary conditions
$\phi(0, p)=0$, $ \phi(X, 0)=\phi(1, p) =1$ 
together with the Neumann boundary condition  $\bigl[ \sqrt{1-p}\, \partial_p \phi  (X, p)\bigr]\big|_{p=1} =0$, which can be replaced by
the free boundary condition at $p=1$ as before.  We shall solve the equation  for  the half region of $ X\ge 0$ utilizing the underlying $\mathbb{Z}_2$
symmetry.
Below we shall pay a particular attention to the horizon profile,  $\phi(X, 1)$, of  our scalar field to see how the horizon is colored by the scalar hair.

Next we would like to consider a bag-like configuration 
as an application of the Janus deformation of the black brane background. For this, we introduce a boundary radial coordinate $r$ defined by
$r=\sqrt{x_1^2+x_2^2 + x_3^2}$ and impose the boundary condition $\phi(r, 0)=-\gamma \, \theta(R-r)$ where $\theta(x)$ denoting the 
Heaviside step function. This boundary condition describes a bag-like model where the YM coupling  $g^2_{YM}= 4\pi \, e^{{\phi(r,0)}+\phi_0}$ 
in the region of $r \le R$
 becomes weaker than the one outside the bag. Later we shall argue that hadrons can be realized by a fundamental (QCD) string corresponding to
a Wilson line connecting  quark to anti-quark in the YM theory side. Redefining $S= r \, \phi$ with $X=\tanh r$, one finds the deformation is 
described by 
\be\label{eqBag}
(1-X^2) \partial_X \Bigl[ (1-X^2) \partial_X  S \Bigr] + 4p(1-p^2)\partial^2_p  S-4 (1+p^2) \partial_p S =0
\ee
which is precisely the same form as (\ref{5Deq}). Setting $\gamma=1$ by utilizing the linearity of the problem, 
we have the boundary conditions $S(0, p)=S(1, p)=0$, $ S(X, 0)= -{\rm arctanh} X \, \theta (\tanh R -X)$ together with the 
free boundary condition 
at $p=1$. In this configuration, we have one more scale given by the bag size $R$ in addition to the temperature $T$ of the black brane. 
Thus the underlying scale symmetry is broken leading to a potentially richer dynamics which may depend on the parameter $RT$.

\section{3D Black Janus by Monte Carlo}

In this section, we shall {give a detailed description of the MC method to solve the scalar equation \eqref{3dEq}}, which has a nice geometrical interpretation in the gravity problem as discussed in the introduction.
{We begin with discretizing (\ref{3dEq})} with boundary conditions $\phi(0,p)=0$, $\phi(1,p)=\phi(X,0)=1$ 
and the free boundary condition at the horizon. 
The finite difference approximations (FDA) based on the Taylor expansion to 
the derivatives in (\ref{3dEq}) are given by

\begin{align}
&\partial_X \phi \to \frac{\phi_{i+1,j} - \phi_{i-1,j}  }{2~\Delta X}~,~\partial_X^2 \phi \to \frac{\phi_{i+1,j} - 2\phi_{i,j} + \phi_{i-1,j}  }{{\Delta X}^2}~,~\\
&\partial_p \phi \to \frac{\phi_{i,j+1} - \phi_{i,j-1}  }{2~\Delta p}~,~\partial_p^2 \phi \to \frac{\phi_{i,j+1} - 2\phi_{i,j} + \phi_{i,j-1}  }{{\Delta p}^2}~,
\end{align}
where $i$ and $j$ correspond to the $X$ and $p$ coordinates of site $(i,j)$ for 
a two-dimensional rectangular lattice with the grid spacing of $\Delta X=1/N_X$ 
and $\Delta p=1/N_p$, respectively.
Then the FDA representation for the 3D Janus equation (\ref{3dEq}) can be
written as
\begin{align}\label{eqDiff01}
\phi_{i,j} = P(i+1,j)~\phi_{i+1,j}
	+ P(i-1,j)~\phi_{i-1,j}
	+ P(i,j+1)~\phi_{i,j+1}
        + P(i,j-1)~\phi_{i,j-1},
\end{align}
where
\begin{align} \label{pi1j}
&P(i\pm1,j) = \frac{{(1-X_i^2)^2}/{\Delta X}^2 
	\mp X_i(1-X_i)^2/{\Delta X}}
	{D_{i,j}} \\
&P(i,j\pm1) = \frac{4p_j(1-p_j)/{\Delta p}^2 
	\mp 2p_j/{\Delta p}}
	{D_{i,j}},\label{pij1}
\end{align}
and $D_{i,j}=2(1-X_i^2)^2/{\Delta X}^2 + 8p_j(1-p_j)/{\Delta p}^2$.
This form allows us to interpret the coefficients $P(i\pm1,j)$ and 
$P(i,j\pm1)$ as the probabilities that a random walker
moves to neighboring sites along $X$ and $p$ directions.
{Since they are determined by the metric, which is the black brane
metric \eqref{bbmetric} in this case, it is clear that they carry
the information of the geometry of the problem.}
Note that the summation of the probabilities gives unity as it should be.

Reformulating the problem in this way, we are naturally led to
solve the equation using the MC method. The 
solutions $\phi_{i,j}$ are determined through samples of random walkers with 
the probabilities of \eqref{pi1j} and \eqref{pij1}. 
MC updates consist of 
interchange of positions of a random walker between a given site and one of its 
four neighboring sites. Updates are sequentially attempted along $X$ and $p$ 
directions until the walker arrives at a boundary.
If an update is accepted according to the 
following hopping rule: 
\begin{align}
&\xi \in (0,P(i+1,j)]&:&~~\text{increase}~~i~~\text{by 1}   \nonumber\\
&\xi \in (P(i+1,j),P(i+1,j)+P(i-1,j)]&:&~~\text{decrease}~~i~~\text{by 1}  \nonumber\\
&\xi \in (P(i+1,j)+P(i-1,j),P(i+1,j)+P(i-1,j)+P(i,j+1)]&:&~~\text{increase}~~j~~\text{by 1}   \nonumber\\
&\xi \in (P(i+1,j)+P(i-1,j)+P(i,j+1),1]&:&~~\text{decrease}~~j~~\text{by 1}  \nonumber
\end{align}
a new site is stored in a previous site. Here, $\xi\in(0,1]$ is a uniform 
random number. Denote $n_{(i,j),\hat a}$ as the number of arrivals that a walker 
starting from a site $(i,j)$ reaches to a boundary site $\hat a$ out of total $N_{MC}$ Monte-Carlo arrivals.
The set $\{n_{(i,j),\hat a}\}$ is the ``experimental'' data obtained by the random 
walker after exploring the geometric landscape.
Our measurements of the solutions are then evaluated with the estimator
\begin{align} \label{phiij}
\phi_{i,j} = \sum_{\hat a} \frac{n_{(i,j),\hat a}}{N_{MC}} \phi_{\hat a}
\end{align} 
We perform the same procedure for all $(i,j)$ which consists of one MC block. 
We then repeat this many times for the sake of error estimation and numerical stability.

In fact, MC methods have been one of standard ways to solve PDEs
numerically \cite{MC01}. There are several advantages to using MC simulations.
While direct or iterative numerical approximations for PDEs often suffer from 
requiring an adequate initial guess of a trial solution, the use of MC sampling does 
not require any trial solution and leads to the advantage of straightforward parallel 
computing. The MC method we use can deal with an arbitrary geometry and 
be extended to nonlinear boundary value problems with various boundary conditions.

While most applications have been to Dirichlet boundaries, this approach 
can also be applied to more complicated boundary conditions such as Neumann 
and free boundary conditions where the Dirichlet boundaries are 
determined through sampling to evaluate appropriate $\phi_{\hat a}$'s 
regardless of details of equations. In our problem, we have a free boundary
condition at the horizon that $\phi(X,1)$ is smooth. This condition can
be implemented by requiring that the second order derivatives should be
of the order $\Delta p$ at the horizon, i.e.,
\begin{align} \label{freebc}
  ( \phi_{i,N_p} - 2\phi_{i,N_p-1} + \phi_{i,N_p-2}) 
  - ( \phi_{i,N_p-1} - 2\phi_{i,N_p-2} + \phi_{i,N_p-3})   \sim \mathcal O ( \Delta p^3 )~~,
\end{align}
which can be put to zero in the FDA employed here. Together with 
\eqref{phiij} applied to $j=N_{p}-1, N_{p}-2$ and $N_{p}-3$, \eqref{freebc}
becomes a set of linear equations for $\phi_{i,N_p}$. With $\phi_{i,N_p}$
determined, the boundary reduces effectively to Dirichlet and we can proceed 
to conduct MC estimations for the rest of the sites.

\begin{figure}
\begin{center}
\includegraphics[width=8cm,clip]{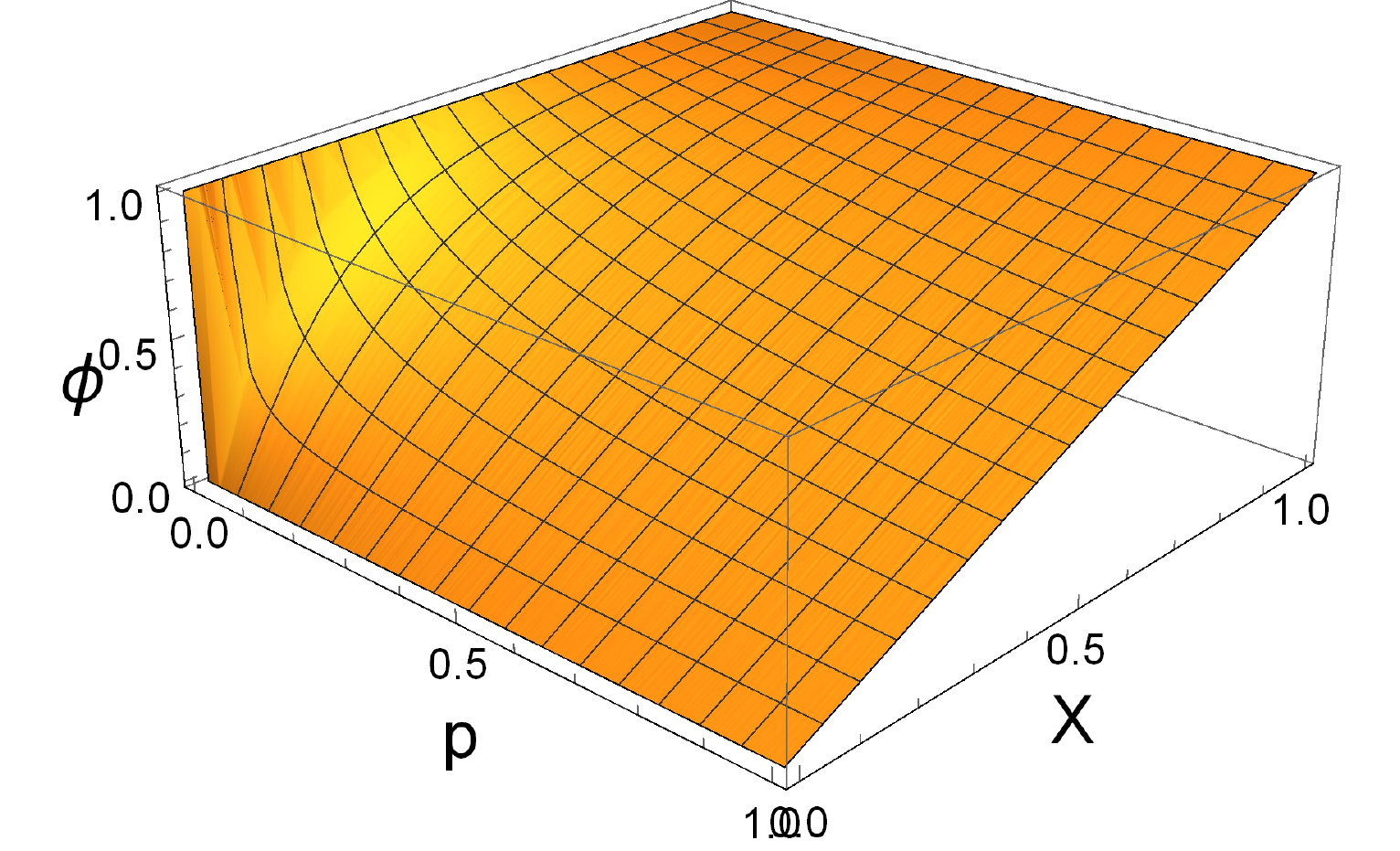}
\includegraphics[width=7cm,clip]{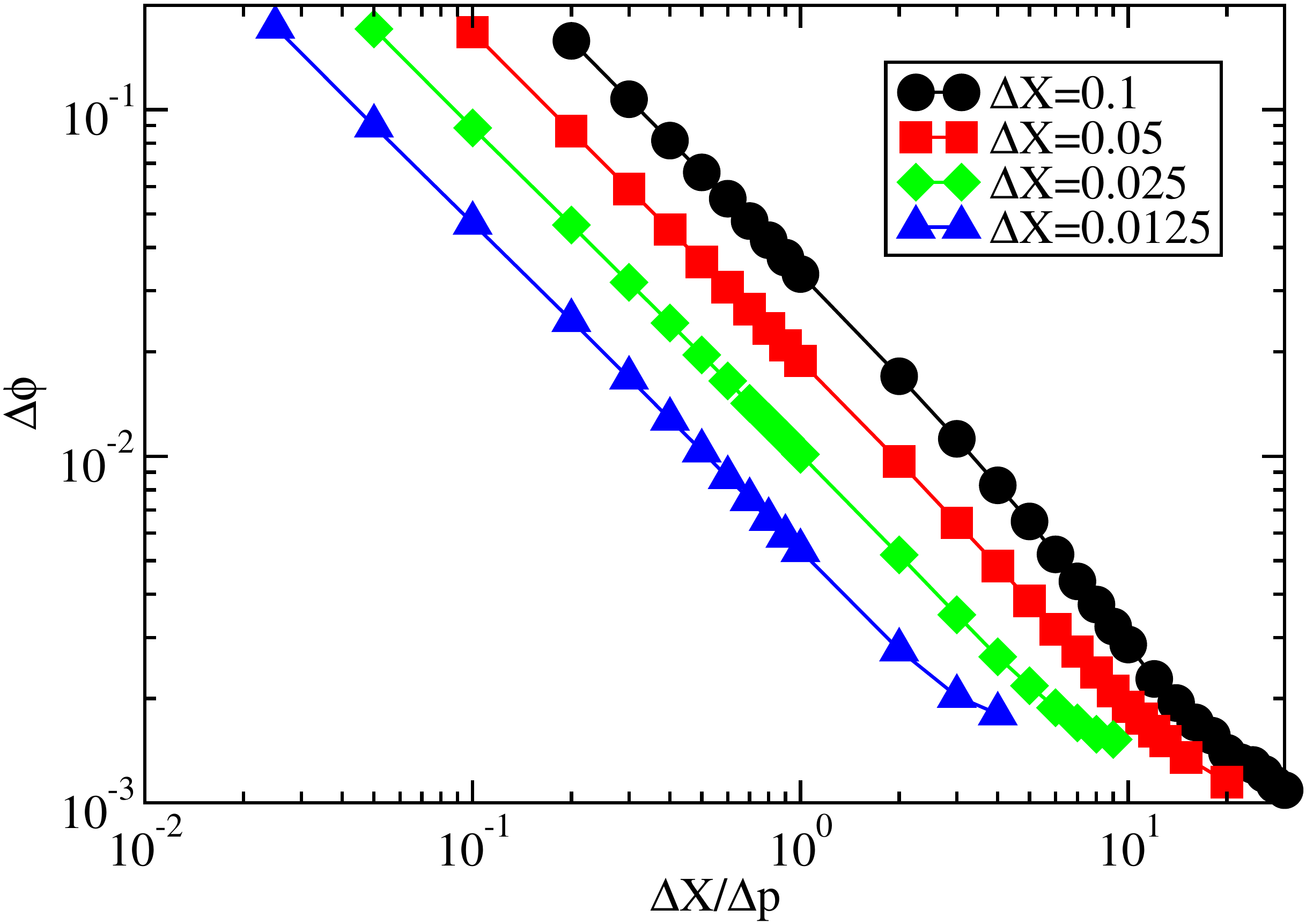}
\end{center}
\caption{
\label{fig01}
Numerical solution of the 3D scalar field is depicted on the left. Relative error is depicted on the right as a function of ${\Delta X}/{\Delta p}$
with the sampling of 500 blocks. Each block contains $N_{MC}$=5000 random walkers.
Circles, squares, diamonds and triangles are for ${\Delta X}$=0.1, 0.05, 
0.025 and 0.0125, respectively. 
}
\end{figure}
  
Figure~1 shows the numerical solution for the 3D scalar profile 
which is indistinguishable from the exact solution \eqref{exact3Dsol} within the resolution of our figure. 
In particular, the profile at the horizon 
correctly behaves as $\phi(X,1)=X$, which confirms the effectiveness 
of the free boundary condition \eqref{freebc} adopted here.
On the right, we also plot the average of the relative errors of the numerical 
solution,
$\Delta\phi\equiv \frac{1}{N_X N_p} \sum_{i,j}
\langle |(\phi_{i,j}^{\rm MC}-\phi_{i,j}^{\rm exact})/\phi_{i,j}^{\rm exact}| \rangle $
over 500 MC blocks, for various lattice sizes. The plot shows that the MC method produces
fairly good numerical solutions nicely converging to the exact solution. 
Note that the error can be reduced by asymmetric
discretization. Indeed, it decreases as a power law in $\Delta X/ \Delta p$. 
This is because of the singular nature of the solution at the origin where
the field value abruptly changes in the $p$ direction as is evident from
the boundary conditions $\phi(0,p)=0$ and $\phi(X,0)=1$.

Closing this section, we remark that we have also tried other numerical 
methods such as the relaxation method and the pseudo-spectral method, 
and obtained solutions with accuracies comparable to MC solutions.

\section{5D Black Janus}

In five dimensions, we consider two kinds of the interfaces. One is planar and the other spherical.

\vskip 0.5cm
\noindent\underline{\it{Black Janus with Planar Interface} }

For the planar interface sitting at $x_1=0$, the corresponding scalar field is governed by
(\ref{5Deq}). 
In the FDA of the previous section, we obtain the same form of difference equation 
(\ref{eqDiff01}) with the following probabilities {determined by the
black brane geometry},        
\begin{align}
&P(i\pm1,j) = \frac{{(1-X_i^2)^2}/{\Delta X}^2
        \mp X_i(1-X_i)^2/{\Delta X}}
        {D_{i,j}} \label{pp1}  \\
&P(i,j\pm1) = \frac{4p_j(1-p_j^2)/{\Delta p}^2
        \mp 2(1+p_j^2)/{\Delta p}}
        {D_{i,j}},
\label{pp2}
\end{align}
where $D_{i,j}=2(1-X_i^2)^2/{\Delta X}^2 + 8p_j(1-p_j^2)/{\Delta p}^2$.

\begin{figure}
\begin{center}
\includegraphics[width=7cm,clip]{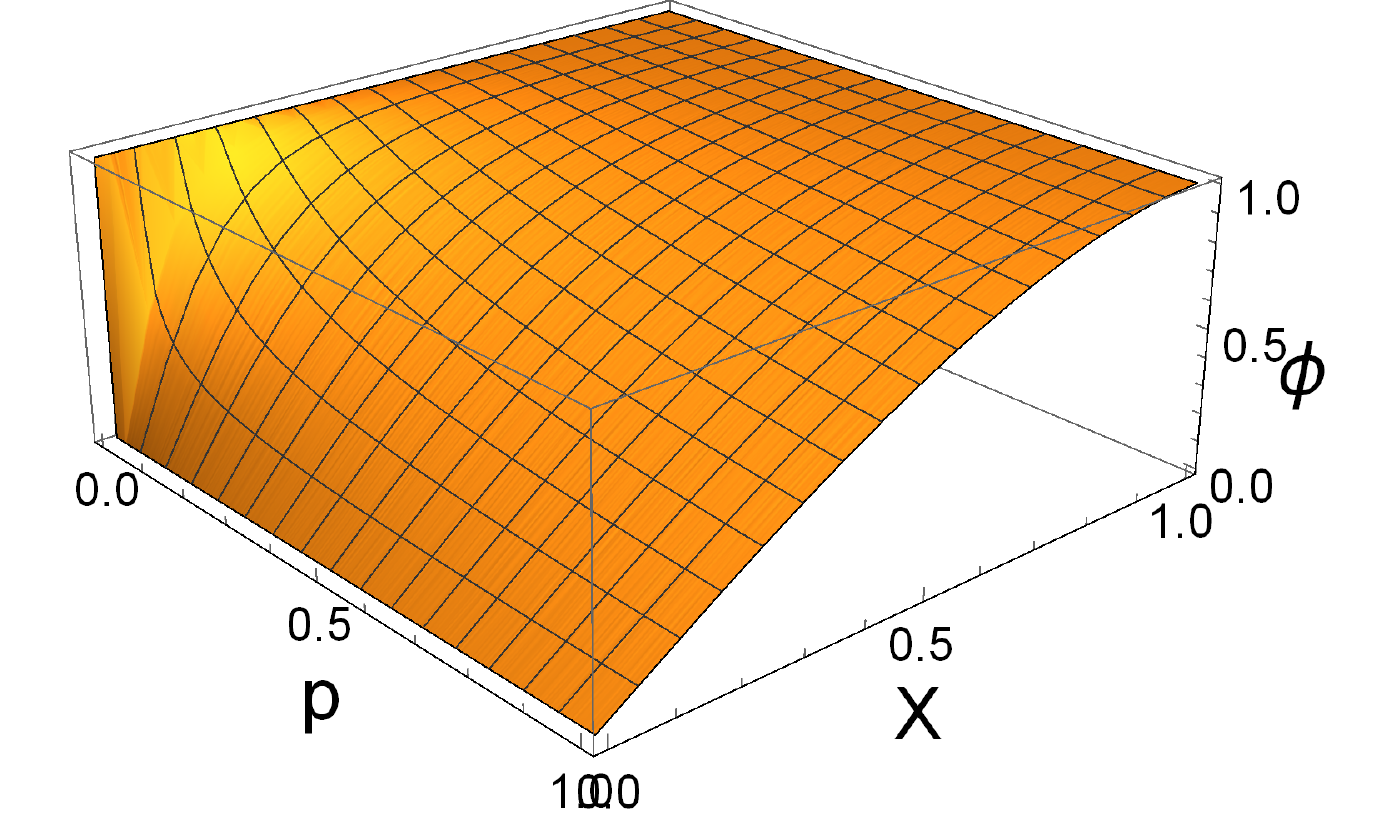}~
\includegraphics[width=6cm,clip]{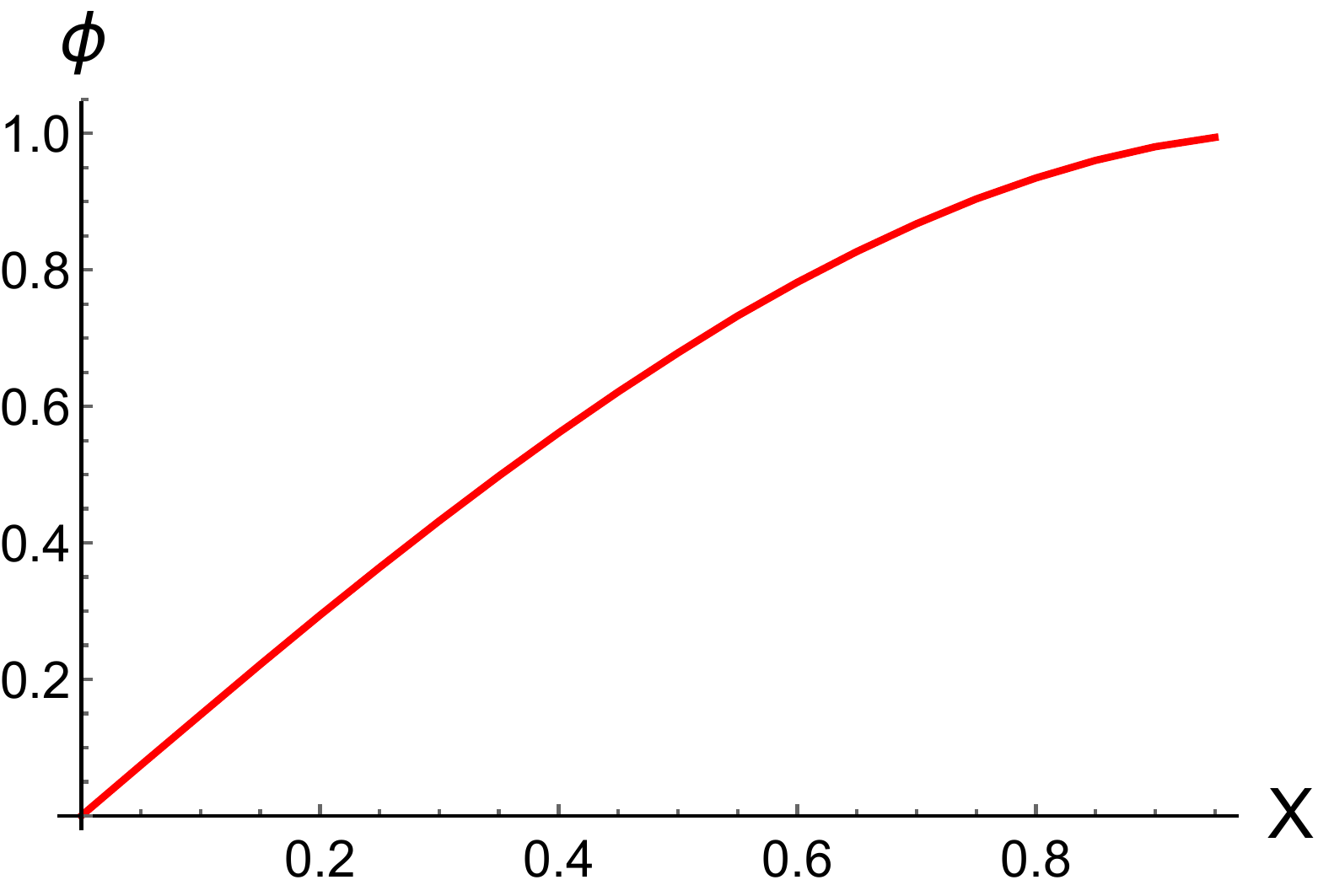}
\end{center}
\caption{
\label{fig02}
Numerical solution of the Janus scalar 
field in 5D with a planar interface is depicted on the left.  Corresponding horizon profile $\phi(X,1)$ is depicted on the right.
}
\end{figure}

The MC solution and the profile at the horizon is presented in 
Fig.~\ref{fig02}. In this case, we do not have the exact solution for comparison, but 
 have checked that the MC solution shows a converging behavior 
as we increase the number of MC blocks. The overall shape of the solution
is similar to Fig.~1. This is natural because,
at the three sides, Dirichlet boundary conditions are imposed with the same 
boundary values as the three-dimensional case. At the horizon, however,
the field is allowed to take any finite value and hence shows a nonlinear
profile in Fig.~2 in contrast with the previous case.

\vskip 0.5cm  
\noindent\underline{\it{Bag-like Model and String Dynamics} }

Here we shall solve the equation  (\ref{eqBag}), which is for the scalar  profile involving the 
spherical interface. It is formally the same as (\ref{5Deq}) and, thus, its MC probabilities from ($i,j$) to its 
neighboring sites are given precisely by (\ref{pp1}) and (\ref{pp2}). The corresponding boundary conditions 
are specified in section \ref{Sec2}. Since the probabilities are the same as those of the planar case, 
we {simply} run the same MC code and then
insert new boundary data into (\ref{phiij}) replacing the old ones. 

\begin{figure}[ht!]
\begin{center}
\includegraphics[width=6cm,clip]{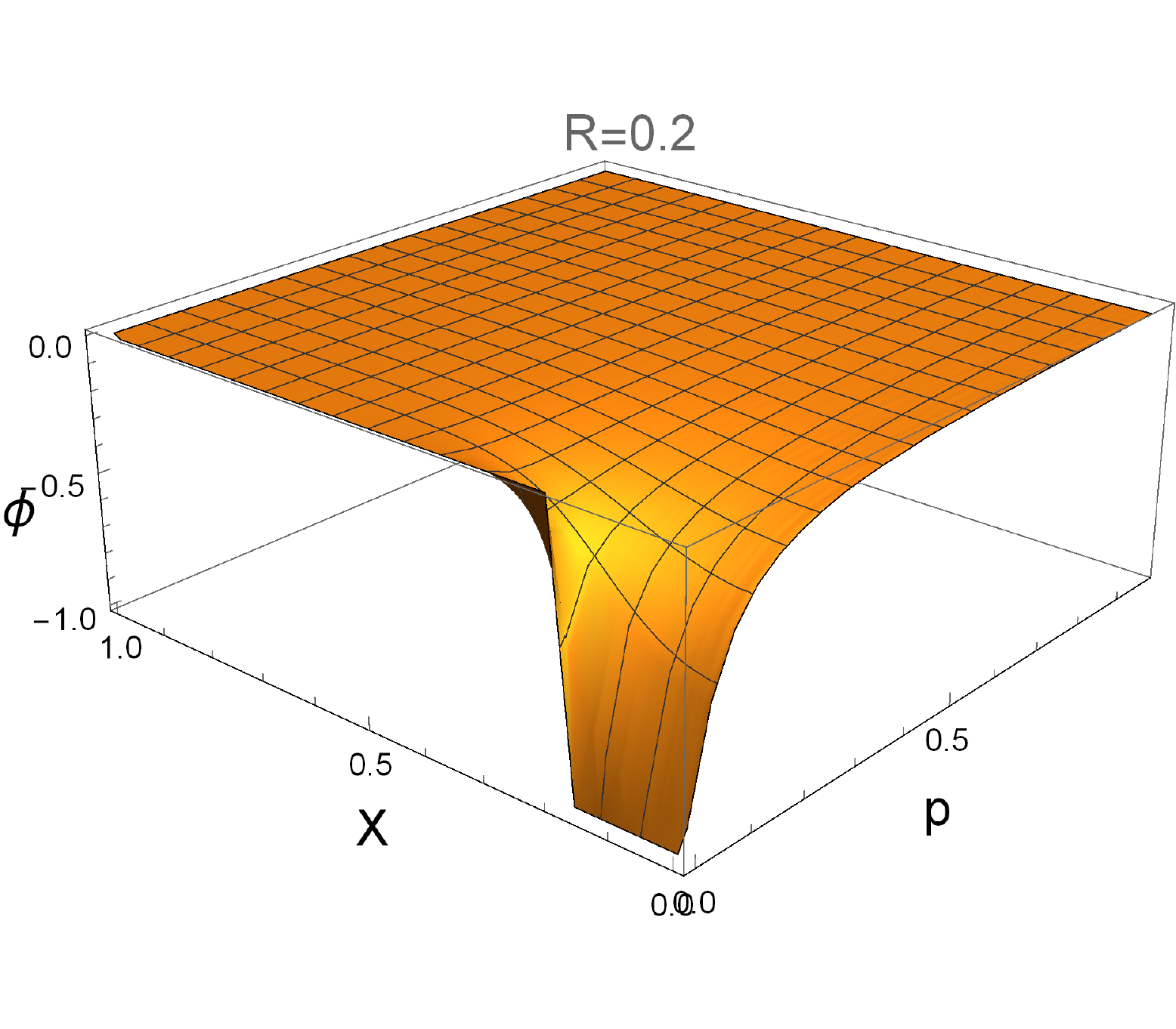}~~
\includegraphics[width=6cm,clip]{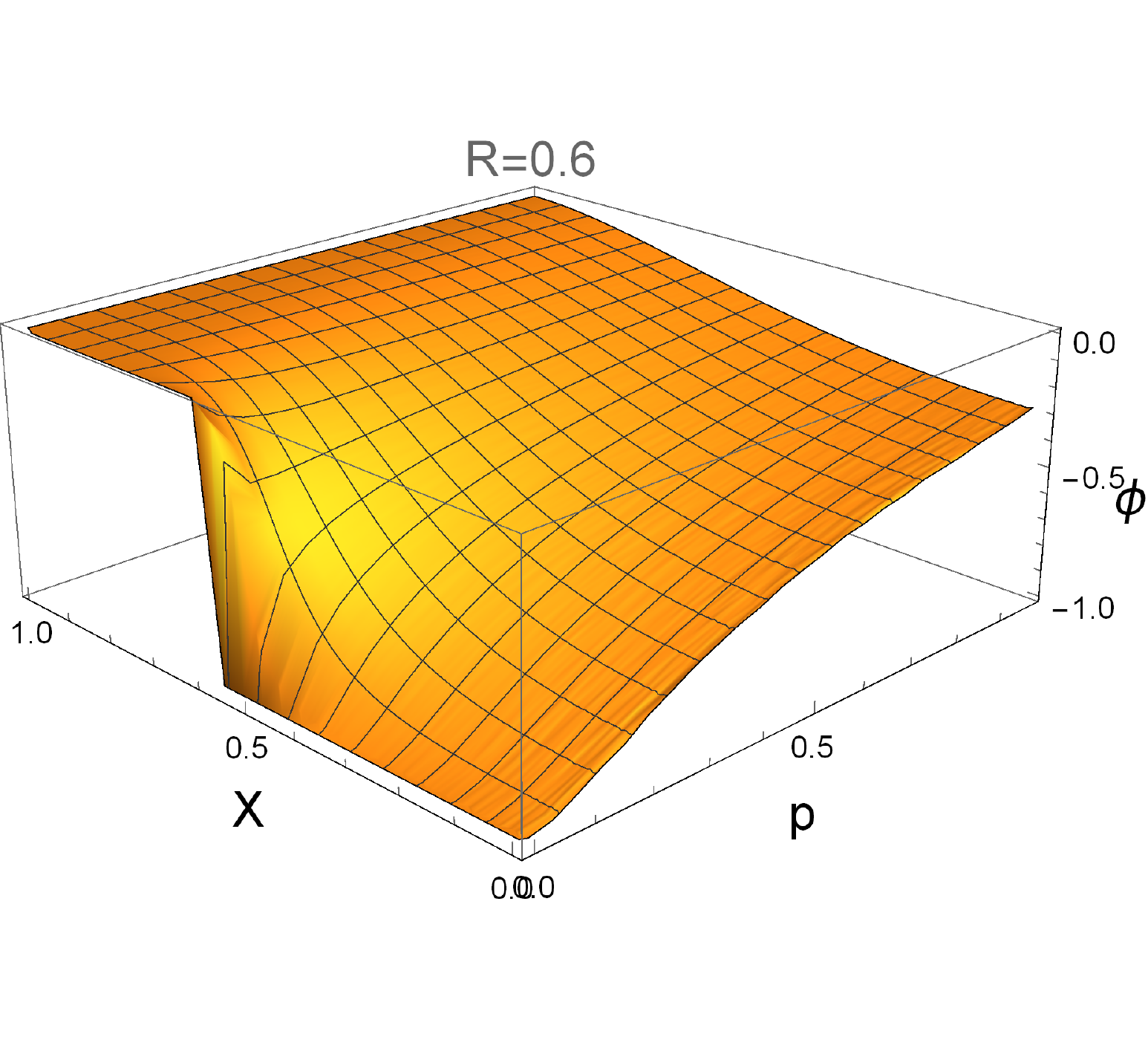}\\
\includegraphics[width=6cm,clip]{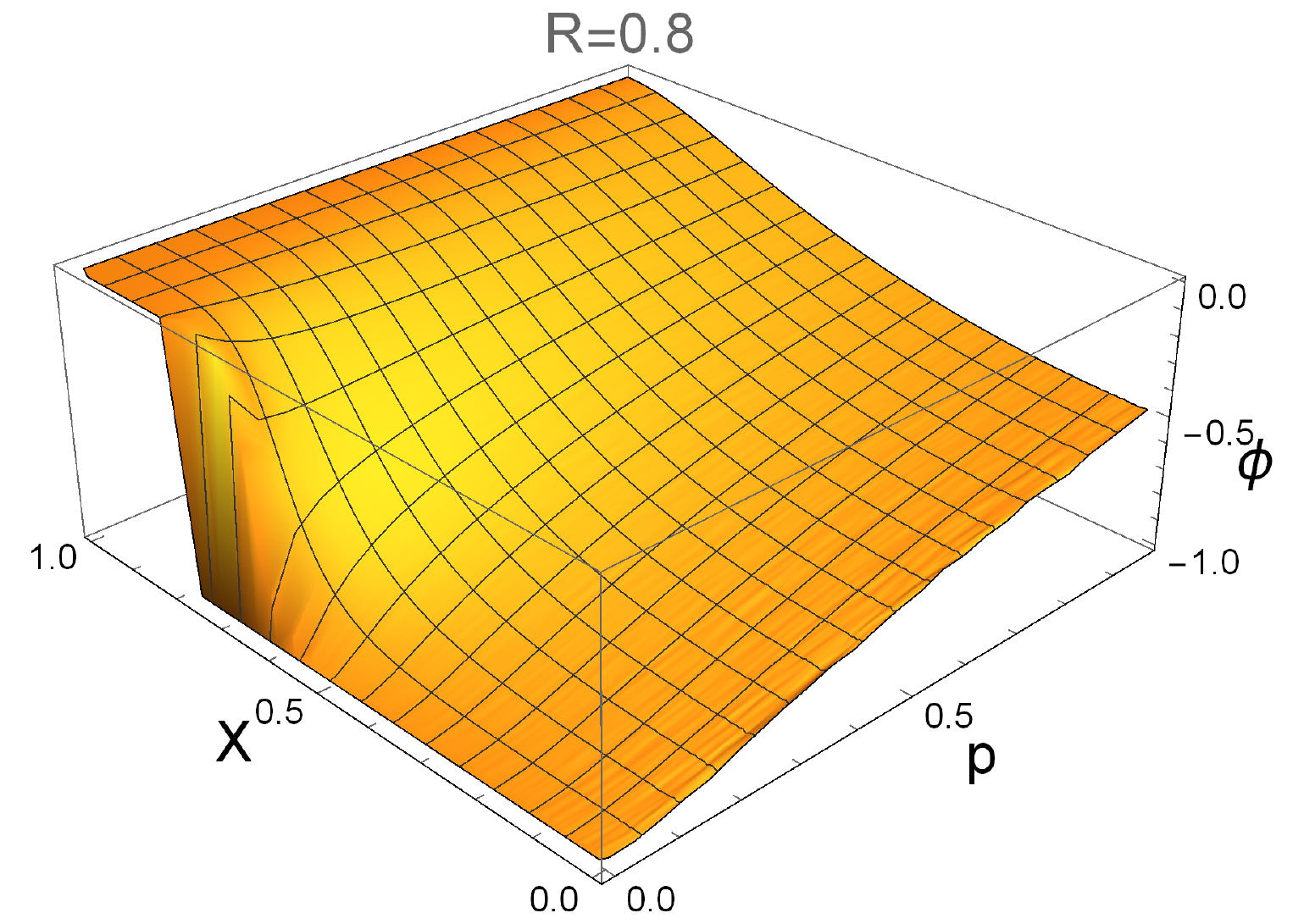}~
\includegraphics[width=6cm,clip]{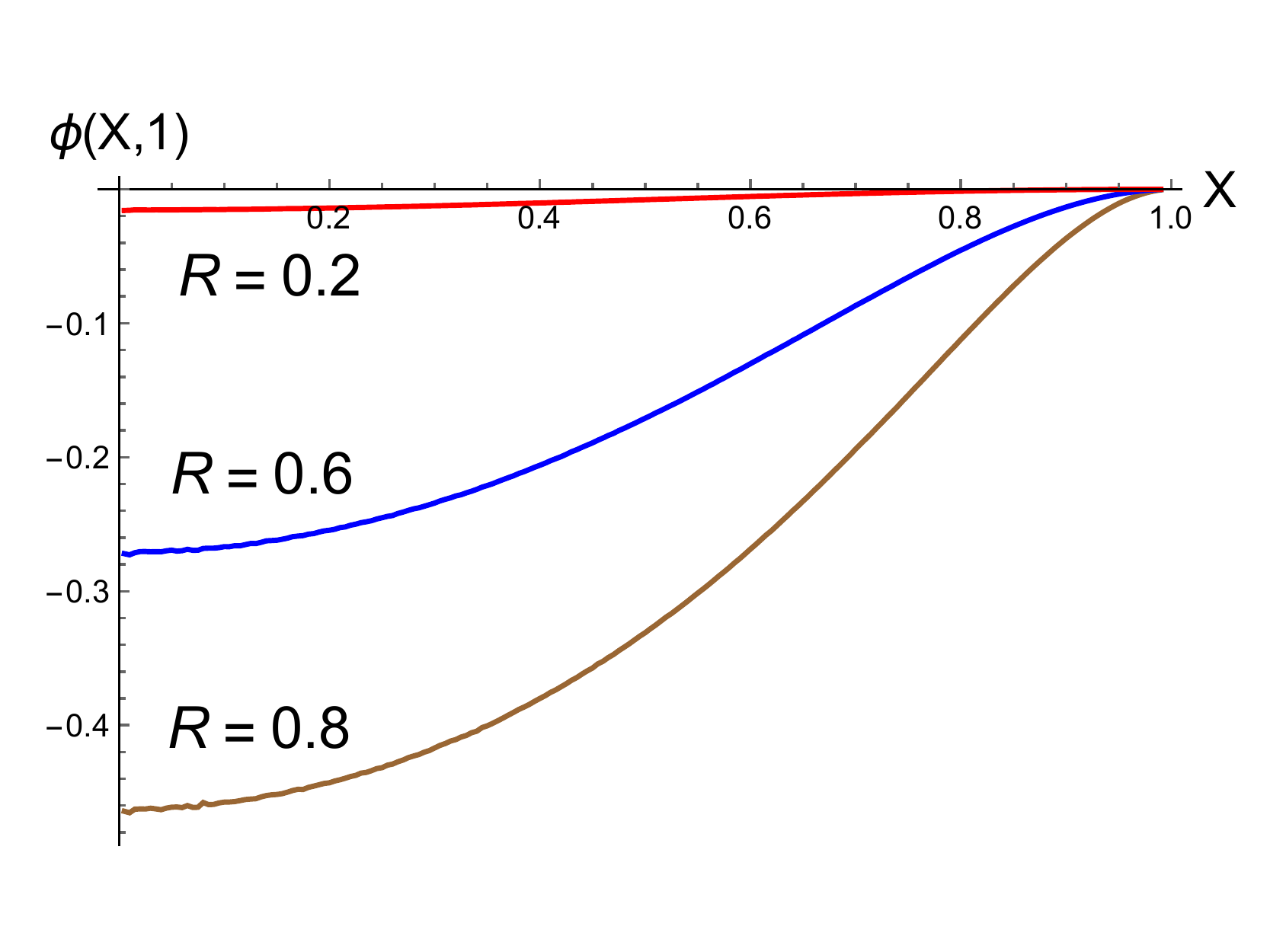}
\end{center}
\caption{
\label{fig04}
The first three are numerical solutions of 5D scalar field with the interface radii R=0.2, 0.6, and 0.8, respectively.
The last graph shows the corresponding horizon profiles again for R=0.2, 0.6, and 0.8, respectively.
}
\end{figure}

In Fig.~\ref{fig04}, we present the resulting scalar profiles for interface radii $R=0.2, 0.6$, and 0.8, respectively. At the boundary 
of geometry ($p=0$), the sharp change of scalar values merely reflects our choice of bag-like boundary conditions. Outside the interface sphere,
we can see that the scalar values approach zero and the corresponding YM coupling is identified with $g^{out}_{YM} = 
\sqrt{4\pi}  e^{\frac{1}{2}\phi_0}$ whereas $g^{in}_{YM} = 
e^{-\frac{1}{2}\gamma}g^{out}_{YM}$ reinstating our interface coefficient $\gamma$. For large $\gamma$, the coupling inside is 
significantly weaker than the one outside. The horizon profiles of the scalar field are also depicted in Fig.~\ref{fig04}. One can see that 
the sharp change at $p=0$ is now smoothened and the absolute values of scalar at the horizon location $X$ become smaller than that
of the boundary location, 
i.e. $ |\phi(X, 1)| < |\phi(X,0)|$.  Since $RT$ is the only scale-independent parameter of our model,
the larger  $R$ with fixed $T$ can be traded by the larger $T$ with fixed $R$ describing essentially the same physics.

 In our bag-like 
model, we are interested in the setup where one dials the temperature while fixing bag radius to the scale of $1/\Lambda_{QCD}$.
Of course in order to get any quantitative results, one has to consider the full back-reacted geometry, which is beyond the 
scope of our present paper. Here instead we would like to briefly discuss only qualitative aspects of our finite-temperature, bag-like 
model. As a probe, one may introduce a pair of quark and anti-quark connected by a Wilson line operator, which is dual to a bulk 
string  in the geometric side \cite{Maldacena:1998im}. In the heavy limit of quark mass, the string worldsheet ends on the boundary of  bulk geometry. By introducing 
 D7 branes in the bulk, the string may end on these D7 branes and the constituent quark masses  can be made finite \cite{Karch:2002sh}. 
In any cases the    
worldsheet dynamics is described by the following Lagrangian \cite{Bak:2003jk},
\begin{align}
L= - \frac{1}{2\pi} \int d\sigma_1 e^{\phi/2} \sqrt{ - \det \partial_\alpha X^a \partial_\beta X^b  g_{ab} }~~,
\end{align}
where $\alpha,\beta=0,1$ are for the worldsheet coordinates ($\sigma_0$, $\sigma_1$) and 
$X^a(\sigma_0,\sigma_1)$ describe the embedding of worldsheet to the bulk spacetime. The crucial part for us is that the effective string tension is 
proportional to $ e^{\phi/2}$, so that the region of smaller effective tension is dynamically preferred. Thus in our bag-like model, 
the quark-antiquark pair  inside the bag is dynamically preferred which is in accordance with the main idea of the MIT-bag model\footnote{However, 
unlike the MIT-bag model, our bag-like model is not confining outside the bag, since the underlying ${\cal N}=4$ SYM theory is non-confining, 
and instead its coupling gets stronger whose strength one may adjust freely.}.
Next to see the finite temperature effect, we note that, as the temperature gets larger, the effective bag radius $RT$ gets larger.
The Debye screening of quark charges will occur within the bag, where effective separation of quark-antiquark exceeds the 
Debye length scale \cite{Rey:1998bq}. In this screening phase, the configuration of two disjoint strings (emanated  respectively  from quark and anti-quark)  
ending on the horizon is preferred to that of connected one. Eventually the bag will melt, which  will lead to the YM plasma phase at high 
enough temperatures, whose details are beyond scope of the present paper.

\section{Discussion}

In this note, we have found numerical solutions for linearized scalar equations in three and five dimensions with various Janus boundary conditions. By comparing our numerical result with the known exact solution in three dimensions, we found a good agreement confirming validity of our geometric MC method. In five dimensions, numerical solutions with two different Janus boundary conditions (one with a planar shape and the other with a spherical shape) have been obtained. 
In our bag-like model with a spherical interface, we argued that the quark-anti-quark pair is dynamically preferred inside the bag as in the original MIT-bag model.
We also discussed temperature dependence of the system as an application of numerical result.


Even though only linearized scalar equations are studied in this note, our method may be 
applied to nonlinear problems 
including full back reaction of the gravity sector. 
Once the back reaction is included, one may study  various quantitative aspects of our 5D system including spectrum of mesons and their melting at finite temperature. 
Further studies are required in these directions.

\section*{Acknowledgement}
D.B. was
supported in part by supported in part by 2016 Research Fund of University of Seoul. 
K.K. was supported by 
NRF  
Grant 
2015R1D1A1A01058220.
H.M. was supported by the 2015 sabbatical-year-research-grant of University of Seoul.

\end{document}